\documentclass[sigconf, nonacm]{acmart}
\setcopyright{acmlicensed}
\copyrightyear{2025}
\acmYear{2025}
\setcopyright{rightsretained}
\acmConference[RecSys '25]{Proceedings of the Nineteenth ACM Conference on Recommender Systems}{September 22--26, 2025}{Prague, Czech Republic}
\acmBooktitle{Proceedings of the Nineteenth ACM Conference on Recommender Systems (RecSys '25), September 22--26, 2025, Prague, Czech Republic}\acmDOI{10.1145/3705328.3759311}
\acmISBN{979-8-4007-1364-4/2025/09}

\usepackage{graphicx} 

\title{Balancing Accuracy and Novelty with Sub-Item Popularity}

\author{Chiara Mallamaci}
\email{chiaramallamaci043@gmail.com}
\orcid{0009-0003-2663-3068}
\affiliation{%
  \institution{Politecnico di Bari}
  \city{Bari}
  \country{Italy}
}

\author{Aleksandr V. Petrov}
\email{spetrov@tripadvisor.com}
\orcid{0000-0002-0911-3605}
\affiliation{%
  \institution{Viator, TripAdvisor\\University of Glasgow}
  \city{Glasgow}
  \country{United Kingdom}
}

\author{Alberto Carlo Maria Mancino}
\email{alberto.mancino@poliba.it}
\orcid{0000-0001-8027-9475}
\affiliation{%
  \institution{Politecnico di Bari}
  \city{Bari}
  \country{Italy}
}

\author{Vito Walter Anelli}
\email{vitowalter.anelli@poliba.it}
\orcid{0000-0002-5567-4307}
\affiliation{%
  \institution{Politecnico di Bari}
  \city{Bari}
  \country{Italy}
}

\author{Tommaso Di Noia}
\email{tommaso.dinoia@poliba.it}
\orcid{0000-0002-0939-5462}
\affiliation{%
  \institution{Politecnico di Bari}
  \city{Bari}
  \country{Italy}
}

\author{Craig Macdonald}
\email{craig.macdonald@glasgow.ac.uk}
\orcid{0000-0003-3143-279X}
\affiliation{%
  \institution{University of Glasgow}
  \city{Glasgow}
  \country{United Kingdom}
}

\renewcommand{\shortauthors}{Chiara Mallamaci et al.}



\usepackage{booktabs}  
\usepackage{stfloats}
\usepackage[subrefformat=parens,labelformat=parens]{subcaption}
\usepackage{multirow}     
\usepackage{cleveref}

\definecolor{redline}{RGB}{207, 33, 39} 
\definecolor{blueline}{RGB}{30, 109, 168} 
\definecolor{greenline}{RGB}{44, 148, 49}

\begin{document}
\begin{abstract}
In the realm of music recommendation, sequential recommenders have shown promise in capturing the dynamic nature of music consumption. A key characteristic of this domain is repetitive listening, where users frequently replay familiar tracks. 
To capture these repetition patterns, recent research has introduced Personalised Popularity Scores (PPS), which quantify user-specific preferences based on historical frequency.
While PPS enhances relevance in recommendation, it often reinforces already-known content, limiting the system’s ability to surface novel or serendipitous items—key elements for fostering long-term user engagement and satisfaction.
To address this limitation, we build upon RecJPQ, a Transformer-based framework initially developed to improve scalability in large-item catalogues through sub-item decomposition. We repurpose RecJPQ’s sub-item architecture to model personalised popularity at a finer granularity. This allows us to capture shared repetition patterns across sub-embeddings—latent structures not accessible through item-level popularity alone.
We propose a novel integration of sub-ID-level personalised popularity within the RecJPQ framework, enabling explicit control over the trade-off between accuracy and personalised novelty.
Our sub-ID-level PPS method (sPPS) consistently outperforms item-level PPS by achieving significantly higher personalised novelty without compromising recommendation accuracy. 
Code and experiments are publicly available at \url{https://github.com/sisinflab/Sub-id-Popularity}.

\end{abstract}

\maketitle

\section{Introduction}
Sequential recommendation aims to predict a user’s future interests by modelling the sequence of their past interactions. Unlike traditional collaborative filtering approaches that treat user history as an unordered set, sequential models leverage the temporal ordering of interactions to capture evolving user preferences~\cite{DBLP:conf/ijcai/WangHWCSO19, DBLP:journals/csur/QuadranaCJ18}.               
Recent advances in sequential recommendation have centred around Transformer-based architectures, drawing inspiration from their success in natural language processing. In these models, tracks or items are akin to tokens in a sentence, and the goal is analogous to masked- or next-token prediction. Notable examples include SASRec~\cite{DBLP:conf/www/ChoHKY21}, its recent adaption gSASRec~\cite{DBLP:conf/recsys/PetrovM23} and BERT4Rec~\cite{DBLP:conf/cikm/SunLWPLOJ19} which are commonly deployed~\cite{DBLP:conf/sigir/TranSSH23, DBLP:conf/cikm/ZhouWZZWZWW20,DBLP:journals/ijcrowdsci/ZhangC0LW21}.
Despite their effectiveness, these models often struggle with sequences characterised by repeated consumption patterns, a crucial aspect in the music recommendation domain. Specifically, unlike many other domains, music consumption is highly repetitive, usually involves passive engagement, and must deal with vast catalogues with sparse individual user histories~\cite{DBLP:books/daglib/0025137, DBLP:conf/recsys/TranSSH24}.
One of the strategies adopted to address this issue involves taking into account item popularity within individual sequences. Popularity is, in fact, one of the simplest and most effective baselines in recommendation~\cite{DBLP:conf/ecir/AnelliNSRT19, 10.1609/aaai.v33i01.33014806}. Modelling it at the level of specific sequences provides a simple yet effective way to capture repetitive consumption patterns.

One such approach proposed augmenting Transformer-based architectures with user-specific recurrence signals, known as Personalised Popularity Scores (PPS)~\cite{DBLP:conf/recsys/AbbattistaANMP24}.
Unlike methods based on global popularity, PPS provides a personalised popularity signal that reflects habitual listening behaviour, significantly improving predictive performance in repetition-heavy settings like music streaming. 
However, heavy reliance on such scores can result in recommendations that reinforce previously seen content, limiting users’ exposure to new or unexpected items. This affects the serendipity of recommendations — i.e., the likelihood of surfacing unexpected yet relevant items — which is widely recognised as crucial for maintaining user satisfaction and promoting long-term engagement~\cite{10.1145/2926720, 10.1145/1864708.1864761, kotkov2016survey,10135}.
%
To address this problem, we take inspiration from RecJPQ~\cite{DBLP:conf/wsdm/PetrovM24}, a Transformer-based framework, originally introduced to address scalability in large-item catalogues through sub-item decomposition. 
In RecJPQ, each item is represented as a combination of shared sub-embeddings, which significantly reduces the number of parameters and the overall computational cost.
Sub-item decomposition has recently gained traction in the recommendation field
~\cite{DBLP:conf/nips/RajputMSKVHHT0S23, DBLP:conf/recsys/SinghV0KSZHHWTC24}. 
These approaches represent items as tuples of quantised semantic IDs, achieving state-of-the-art performance and improving the generalisation capabilities of recommender systems.
While our work specifically focuses on RecJPQ's sub-IDs, we expect the approach to generalise well to other sub-ID–based methods.
Although RecJPQ was primarily developed to improve scalability and efficiency in large-item catalogues~\cite{DBLP:conf/recsys/PetrovMT24, DBLP:conf/wsdm/PetrovM24}, we repurpose its sub-item architecture to address a different problem: enhancing personalisation in sequential recommendation.
We posit that RecJPQ's quantised semantic IDs may also be utilised to compute more fine-grained and personalised popularity signals, thereby increasing user-specific novelty.
By computing personalised popularity scores at the sub-id level and integrating them into sequential models with varying levels of intensity, our study shows that it is possible to tune the balance between personalisation and generalisation.  
Framing the task as an accuracy–novelty trade-off, we compare the effects of injecting personalised popularity at both the item and sub-id levels. 

Our contributions are twofold: 
(1) we propose a novel method for incorporating personalised popularity into sequential recommendation by applying PPS at both item and sub-item levels via RecJPQ’s decomposition strategy;
(2) we conduct an empirical evaluation of the approach.
Results demonstrate that modelling sub-id popularity yields a fine-grained personalisation signal that captures user repetition patterns, reveals structural similarities between items, and enables explicit control over the trade-off between accuracy and personalised novelty.

\begin{figure}[t]
  \centering
  \includegraphics[width=0.8\linewidth]{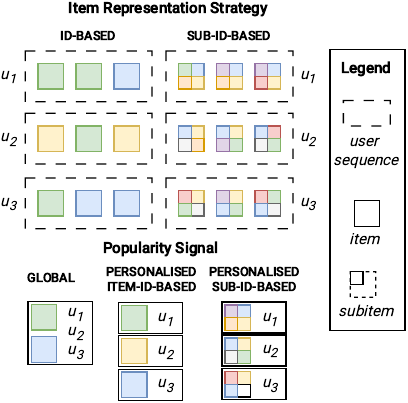}
  \caption{Different approaches for computing item popularity, based on whether the signal is ID-based (global or personalised) or sub-ID-based (personalised).}
  \label{fig:sub_id_pop}
\end{figure}

\section{Methodology}

In this work, we propose a novel personalised popularity-aware approach that enhances sequential recommendation by integrating two complementary user-specific signals:

(1) \textbf{Item-ID Popularity} measures how often a user has listened to a specific track, capturing their tendency to repeat familiar content. Personalised Popularity Scores (PPS)~\cite{DBLP:conf/recsys/AbbattistaANMP24} translate this idea into a simple yet effective signal by assigning higher scores to item-IDs frequently consumed by the user.

(2) \textbf{Sub-ID Popularity} leverages RecJPQ’s compositional item representation, in which each track is encoded as a sequence of sub-identifiers (sub-IDs).
By tracking the appearances of sub-IDs in a user’s history, this signal captures latent repetition, revealing shared components across item-IDs that reflect underlying stylistic or structural patterns (e.g., similar genres or artists). \Cref{fig:sub_id_pop} provides a visual comparison of the popularity signals under the two item representation strategies.

We integrate these two signals with the output of the underlying recommender using a weighted scoring function. Two scalar parameters, $\alpha$ and $\beta$, control the influence of item-ID and sub-ID popularity, respectively.
This integration strategy is model-agnostic and can be applied to any sequential recommender system, as it operates directly at the output scoring level.

\subsection{Preliminaries}
Let $\mathcal{I}$ be the set of items, with $|\mathcal{I}|$ denoting their total number. Each item \(i \in \mathcal{I}\) is associated with a unique \(d\)-dimensional embedding \(\mathbf{w}_i \in \mathbb{R}^d\).
Our method builds on \textbf{RecJPQ}~\cite{DBLP:conf/wsdm/PetrovM24}, which integrates two related quantisation schemes—Product Quantisation (PQ)~\cite{Gray1984VectorQ} and Joint Product Quantisation (JPQ)~\cite{zhan2021jointlyoptimizingqueryencoder}—directly into any sequential recommender.  PQ splits each \(d\)-dimensional item embedding into \(m\) equal-sized segments and assigns each segment to its nearest centroid in a codebook of size \(V\).  JPQ fixes these segment‐to‐centroid assignments before training, allowing end-to-end learning of the much smaller sub‐embedding tables under the recommendation loss, without any post-hoc compression.
Unlike JPQ, RecJPQ assigns sub-ids to items based on an SVD decomposition of the sequence-item interaction matrix.  Under this scheme, each item $i$ is encoded by an \(m\)-tuple of sub-identifiers:
\begin{equation}\label{eq:code}
\mathrm{code}(i) = [z_1^{(i)}, z_2^{(i)}, \dots, z_m^{(i)}],\quad z_k^{(i)} \in \{0, \dots, V{-}1\},\; k = 1,\dots,m ,
\end{equation}
where $z_k^{(i)}$ is the $k^{th}$ code from the codebook of size V, corresponding to a unique sub-embedding.
At inference time, the full $d$-dimensional embedding \(\mathbf{w}_i\) can be reconstructed on the fly by concatenating the $m$ learned sub-embeddings according to $code(i)$, as for RecJPQ. More recent approaches that accelerate the reconstruction process can also be integrated~\cite{DBLP:conf/recsys/PetrovMT24, DBLP:conf/sigir/PetrovMT25}.


\subsection{Modelling Personalised Popularity at the Item Level}

We capture explicit repetition at the item level using Personalised Popularity Scores (PPS)~\cite{DBLP:conf/recsys/AbbattistaANMP24}, which quantify how frequently a user has interacted with each item in their listening history.
Let $c_i$ denote the raw count of item $i$ in the user interaction sequence $\mathcal{S}_u$.
Inspired by~\cite{DBLP:conf/recsys/AbbattistaANMP24}, we compute the PPS value for item $i$, denoted as $\text{PPS}_i$, by first applying a logarithmic transformation with additive smoothing $\epsilon$ to ensure numerical stability:
\begin{equation}
\text{PPS}_i = \log(c_i + \epsilon)\ .
\end{equation}

To make these scores comparable across users, we then apply Z‐score normalisation :
\begin{equation}\label{eq:normalization}
\text{PPS}_i^{\text{std}} = \frac{\text{PPS}_i - \mu_{\text{PPS}}}{\sigma_{\text{PPS}}} ,
\end{equation}
where $\mu_{\text{PPS}}$ and $\sigma_{\text{PPS}}$ are the mean and standard deviation of PPS values over the item set.

\subsection{Modelling Personalised Popularity at the Sub-ID Level}
Instead of treating each item as an atomic vector, RecJPQ encodes it using a sub-ID-based code (see Eq.~\eqref{eq:code}).
By deriving sub-IDs through an SVD-based quantisation, RecJPQ naturally groups items that share similar latent characteristics (e.g., musical genre or artist).

This property is central to our approach: each sub-ID can be interpreted as a compact representation of a meaningful concept, and repeated occurrences of the same sub-IDs in a user’s history reveal implicit preferences for those underlying attributes.
By modelling sub-ID popularity, we aim to reinforce such fine-grained user preferences, enabling more novel and personalised recommendations.

Let $\mathcal{S}_u$ be the interaction sequence of user $u$, for each split $j \in \{1, \dots, m\}$ we define the count of sub-ID $k \in \{0, \dots, V{-}1\}$ for user $u$ as:
\begin{equation}
c_j^{(u)}[k]
\;=\;
\sum_{i \in \mathcal{S}_u} \mathbf{1}\bigl\{ z_j^{(i)} = k \bigr\},
\end{equation}
where $z_j^{(i)}$ denotes the $j$-th sub-ID of the item $i$ in the sequence $\mathcal{S}_u$.

Given a candidate item $i$, whose sub-ID representation is $\mathrm{code}(i) = [z_1^{(i)}, \dots, z_m^{(i)}]$, we represent its user-specific sub-ID counts as:
\begin{equation}
\mathrm{Counts}^{(u)}(i)
\;=\;
\bigl[ c_1^{(u)}[z_1^{(i)}],\, \dots,\, c_m^{(u)}[z_m^{(i)}] \bigr].
\end{equation}

Consistent with PPS, and following a common practice for handling repetition counts~\cite{DBLP:conf/ecir/PetrovTM25}, we apply a logarithmic transformation with additive smoothing to compute the \textbf{\textit{sub-ID-based Personalised Popularity Score (sPPS)}}:
\begin{equation}
\mathrm{sPPS}_i
\;=\;
\sum_{j=1}^m \log\bigl( c_j^{(u)}[z_j^{(i)}] + \epsilon \bigr).
\end{equation}
This transformation smooths the count function, making it less sensitive to large values.

Finally, we standardise this value as in Eq.~\eqref{eq:normalization} to obtain the normalised score $\mathrm{sPPS}_i^{\mathrm{std}}$.

\subsection{Score Function}

We integrate the sequential model with personalised popularity signals by defining a final scoring function that combines \textbf{three components}: (1) the predicted logits from the RecJPQ-version of a sequential recommendation model, (2) the standardised item-level popularity score (PPS), and (3) the standardised sub-ID popularity score (sPPS). The final score for a candidate item $i$ is computed as:
\begin{equation}
\text{logits}_i^{\text{final}} = \gamma \cdot \text{logits}_i^{\text{rec}} + \alpha \cdot \text{PPS}_i^{\text{std}} + \beta \cdot \text{sPPS}_i^{\text{std}} ,
\label{eq:score}
\end{equation}
where $\alpha$ and $\beta$ are scalar weights controlling the influence of item-level and sub-ID popularity, respectively, and $\gamma = 1 - \alpha - \beta$ ensures a convex combination. This formulation enables a flexible trade-off between memorisation (through PPS), generalisation (via sPPS), and the representations learned by the sequential recommender.

\section{Experimental Setup}

\begin{table}
\setlength{\tabcolsep}{4pt}
\caption{Key statistics for the two music datasets after applying a popularity-based sampling method to select 30,000 items (\textit{N}). \textit{Avg len} and \textit{Med len} denote the average and the median number of interactions per user, respectively.}\vspace{-1em}
\footnotesize
\centering
\begin{tabular}{lccccc}
    \toprule
    Dataset & Users & Items & Interactions & Avg. len & Med. len\\ 
    \midrule
    Yandex & 20862 & 30000 & 28408152 & 1362 & 1275\\
    Last.fm-1K & 990 & 30000 & 4990042 & 5040 & 2605\\
    \bottomrule
\end{tabular}\vspace{-1em}
\label{tab:dataset}
\vspace{1em}
\end{table}
 
\noindent \textbf{Research Questions.} Our objective is to empirically investigate the impact of incorporating two distinct personalised popularity signals into RecJPQ-based sequential recommendation models. We structure our investigation around the following research questions:
\begin{enumerate}
    \item[\textbf{RQ1}] What is the effect of integrating item-level personalised popularity (PPS) into the RecJPQ framework on recommendation performance in the music domain?

    \item[\textbf{RQ2}] Compared with PPS, is the sub-ID-based popularity score (sPPS) able to improve the trade-off between accuracy and personalised novelty in music recommendation?
\end{enumerate}
Our experimental setup largely follows the structure of \cite{DBLP:conf/recsys/AbbattistaANMP24}, including the choice of datasets and evaluation method. We publicly share our
code, data, and instructions for replicating our
experiments\footnote{https://github.com/sisinflab/Sub-id-Popularity}.

\vspace{0.5em}
\noindent \textbf{Datasets.}
We conduct the experiments on two music streaming datasets: the Yandex music events\footnote{\url{https://www.kaggle.com/competitions/yandex-music-event-2019-02-16}} and the Last.fm-1K\footnote{\url{http://ocelma.net/MusicRecommendationDataset/lastfm-1K.html}}~\cite{DBLP:books/daglib/0025137}, following the same preprocessing protocol as in \cite{DBLP:conf/recsys/AbbattistaANMP24}.  Table~\ref{tab:dataset} summarises their key statistics.

\vspace{0.5em}
\noindent \textbf{Model.} 
We evaluate our approach with one of the state-of-the-art Transformer-based sequential recommender: \textbf{BERT4Rec}~\cite{DBLP:conf/cikm/SunLWPLOJ19}. 
To enable sub-item decomposition, we integrate RecJPQ~\cite{DBLP:conf/wsdm/PetrovM24} in the aprec~\footnote{\url{https://github.com/asash/BERT4Rec\_repro}} reproducibility framework.
For sub-item decomposition, after evaluating several architectural settings, we found that the best-performing configuration adopts an embedding dimensionality of \( d = 256 \) and \( m = 32 \) codebook splits. This configuration aligns with the findings of the RecJPQ authors~\cite{DBLP:conf/wsdm/PetrovM24}.
We adopt this setup for all the experiments and evaluations presented in the paper.
To analyse how different configurations of personalised popularity signals affect the balance between recommendation accuracy and personalised novelty, in accordance with Eq.~\ref{eq:score}, we fixed $\alpha=0.4$, $\beta=0.4$ at training time, while at inference time we systematically vary \(\alpha\) (for PPS) and \(\beta\) (for sPPS). 
This design allows us to explore a range of personalisation strategies without modifying the model architecture or retraining.

\noindent \textbf{Evaluation Details.} Following the setup from \cite{DBLP:conf/recsys/AbbattistaANMP24} and best practices in sequential recommendation \cite{DBLP:conf/icmcs/ChouYL15,10.1145/3604915.3608839}, we apply a Global Temporal Split strategy and evaluate model performance with Normalised Discounted Cumulative Gain (NDCG)~\cite{DBLP:journals/tois/JarvelinK02}. 
In practice, we sort all user–item interactions by timestamp and reserve the last 10\% for testing. A similar procedure is used to construct a validation set comprising 10\% of the users: 2,048 randomly selected users for the Yandex dataset and 128 for Last.fm-1K. For the NDCG calculation we apply the relevance mapping from~\citet{DBLP:conf/iui/YakuraNG18, DBLP:journals/umuai/YakuraNG22} for Yandex interactions: \textit{like}=2, \textit{play}=1, \textit{skip}=-1, \textit{dislike}=-2.  When a user interacts multiple times with the same track, only the label with the greatest absolute value is retained (so a later like/dislike replaces any prior play/skip). Finally, all negative labels are treated as zero relevance in the NDCG calculation to ensure stability \cite{10.1145/3340531.3412123}.  

To measure personalised novelty, we follow the approach introduced in~\cite{Patel_2023, 10.1145/2043932.2043955}.  
Novelty is defined as:

\begin{equation}
\text{Novelty@K}_u = \frac{1}{K} \sum_{i \in R_u@K} -\log_2 \left( p(i \mid u) \right),
\end{equation}

where \(R_u@K\) denotes the top-\(K\) recommendation list generated for user \(u\), and \(p(i \mid u)\) is the conditional probability that user \(u\) has previously interacted with item \(i\).
For each user, this probability is estimated using relative frequency:  
\(p(i \mid u) = \max\left( \frac{c_i}{\sum_{j} c_j},\ \epsilon \right)\),  
where \(c_i\) is the number of interactions with item \(i\), and the denominator sums over all item interactions in user \(u\)’s history. The \(\epsilon\) term is fixed to $10^{-8}$ to ensure numerical stability caused by taking the logarithm of zero for unseen items.
This formulation emphasises the delivery of unexpected recommendations by penalising items that are already highly familiar to the user.
All evaluations are conducted with a cutoff of 40, a common choice in music recommendation due to users’ tendency to consume content rapidly and interact with longer ranked lists rather than just the top few items.
\section{Results}
\begin{table}[t]
\centering
\caption{Comparison of NDCG@40 at increasing novelty thresholds for PPS and sPPS on Last.fm and Yandex.}\vspace{-1em}
\label{tab:ndcg_novelty_comparison}

\normalsize  
\resizebox{\columnwidth}{!}{%
\begin{tabular}{llcccc}
\toprule
\textbf{Dataset} & \textbf{Method} & \textbf{Novelty $\geq$ 0} & \textbf{Novelty $\geq$ 10} & \textbf{Novelty $\geq$ 12} & \textbf{Novelty $\geq$ 14} \\
\midrule
\multirow{2}{*}{Last.fm} 
    & PPS        & \textbf{0.4159} & 0.3248 &  0.2749   & 0.2749 \\
    & sPPS  & 0.3296          & \textbf{0.3296} & \textbf{0.3016} & \textbf{0.2846} \\
\midrule
\multirow{2}{*}{Yandex}  
    & PPS        & \textbf{0.1566} & 0.1233 & 0.1150 & 0.0983 \\
    & sPPS  & 0.1298          & \textbf{0.1298} & \textbf{0.1242} & \textbf{0.1149} \\
\bottomrule
\end{tabular}
}
\end{table}

\begin{figure}[t]
  \centering
  \begin{subfigure}{0.8\linewidth}
    \centering
    \includegraphics[width=\linewidth]{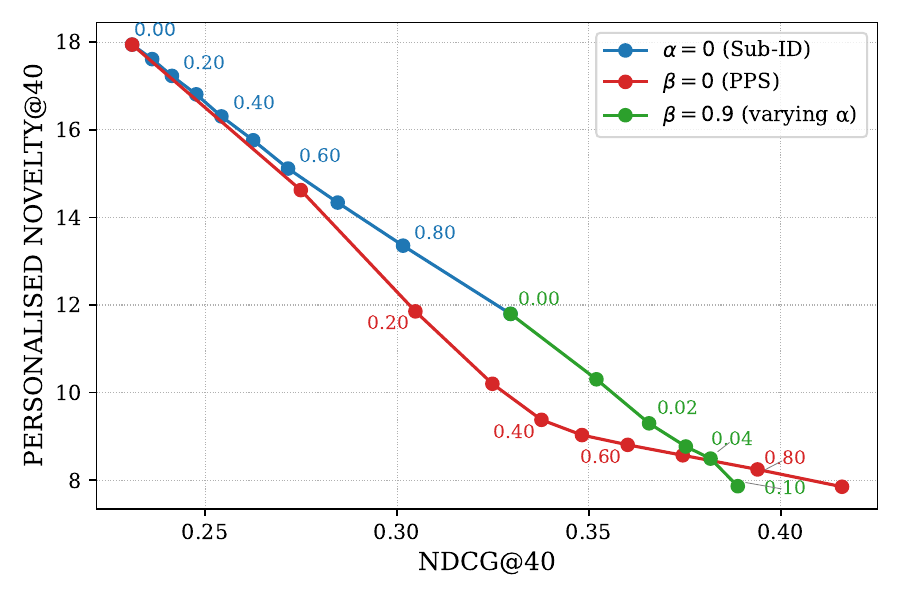}
    \caption{Last.fm dataset}
    \label{fig:tradeoff_lastfm}
  \end{subfigure}
  \vspace{1em}
  \begin{subfigure}{0.8\linewidth}
    \includegraphics[width=\linewidth]{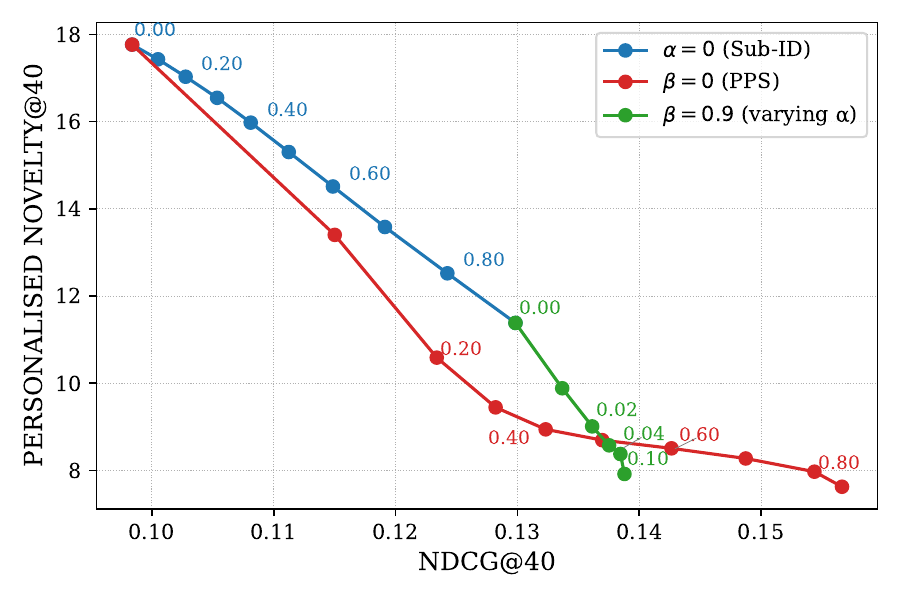}
    \caption{Yandex dataset}
    \label{fig:tradeoff_yandex}
  \end{subfigure}
  \caption{Trade-off between accuracy (NDCG@40) and personalised novelty on both datasets, comparing the effect of applying only PPS (\textcolor{redline}{red}) and sPPS (\textcolor{blueline}{blue}). The \textcolor{greenline}{green line} represents the integration of both signals, with $\beta = 0.9$ fixed and increasing values of $\alpha$.}
  \label{fig:tradeoff_combined}
\end{figure}

Our experimental findings, summarised in Table~\ref{tab:ndcg_novelty_comparison} and visualised in Figure~\ref{fig:tradeoff_combined}, evaluate how item-ID- and sub-ID-level personalised popularity signals affect the trade-off between recommendation accuracy and novelty.

\looseness -1 \noindent \textbf{RQ1.}
\Cref{tab:ndcg_novelty_comparison} and \Cref{fig:tradeoff_combined} (red line) illustrate the effect of incorporating Personalised Popularity Scores (PPS) by varying the parameter \(\alpha\) while keeping \(\beta = 0\). Starting from the baseline configuration (\(\alpha = 0\)), we observe from the plots that gradually increasing \(\alpha\) leads to a consistent improvement in recommendation accuracy across both datasets, as measured by NDCG. This confirms that reinforcing exact-item recurrence through PPS enhances the model’s ability to predict user behaviour.
Focusing solely on accuracy (i.e., for novelty \(\geq 0\)), \Cref{tab:ndcg_novelty_comparison} shows that PPS consistently achieves higher NDCG values than sPPS.
These results confirm that PPS is the stronger performer when the objective is purely predictive accuracy.
However, this gain comes at the cost of reduced personalised novelty. On Last.fm, novelty drops sharply—from 14.62 at \(\alpha = 0.1\) to 7.85 at \(\alpha = 0.9\)—highlighting the model’s growing bias toward previously consumed content. A similar trend is observed on Yandex. These findings reveal a key limitation of item-level popularity modelling: while it enhances accuracy, it constrains the recommender’s ability to support discovery and exploration.

\looseness=-1
\noindent \textbf{RQ2.}
To evaluate the effectiveness of sub-ID popularity modelling, we fix $\alpha=0$ and vary $\beta$, isolating the sub-ID-level signal. As shown in \Cref{tab:ndcg_novelty_comparison} and in \Cref{fig:tradeoff_combined} (blue line), sub-ID modelling consistently outperforms item-level PPS in terms of personalised novelty. For example, for Last.fm (Fig. \ref{fig:tradeoff_lastfm}), at a fixed NDCG@40 of approximately 0.32, the PPS-only model yields a novelty score of about 10, whereas the sub-ID model achieves roughly 12, a 20\% relative improvement at the same accuracy level. These gains are also reflected in \Cref{tab:ndcg_novelty_comparison}: on Last.fm at a novelty threshold of 12, \text{NDCG}@40 climbs from 0.2749 (PPS) to 0.3016 (+9.7\%), and on Yandex at same novelty threshold, \text{NDCG}@40 improves from 0.1150 to 0.1242 (+8.0\%). In short, by capturing structural recurrence at the sub-item level, RecJPQ’s sub-ID popularity signal uncovers latent user preferences beyond exact-item repetition, significantly boosting recommendation novelty without sacrificing accuracy.
To assess the combined effect of the two signals, we evaluate the model when both parameters are active. In particular, we fix $\beta = 0.9$ (strong sub-ID bias) and gradually increase $\alpha$ to introduce item-level PPS. As shown by the green curve in Figure~\ref{fig:tradeoff_combined}, on both datasets, NDCG steadily increases while personalised novelty decreases 
less than in the PPS-only setting. This suggests that the sub-ID signal provides a robust foundation, enabling the integration of PPS to improve accuracy with minimal compromise in novelty.

\section{Conclusion}
This work demonstrates that music recommender systems can substantially benefit from integrating personalised popularity signals. By extending the RecJPQ framework with item-ID and sub-ID level popularity modelling, 
we enable fine-grained personalisation while preserving the efficiency of sub-item decomposition.
Experiments on the Yandex and Last.fm-1K datasets show that modelling sub-ID popularity signals not only improves recommendation novelty without compromising accuracy, but also captures structural recurrence patterns across items--leading to more novel and diverse recommendations compared to item-level PPS.
It also offers explicit control over the accuracy–novelty trade-off, helping the system surface unexpected yet relevant items. 
As future work, we plan to further explore the interplay between item-level and sub-ID-level popularity signals by varying $\alpha$ and $\beta$ parameters directly at training time.
This would provide deeper insights into how layered popularity signals shape the accuracy–novelty trade-off and help identify optimal configurations for different user behaviour profiles.

\bibliographystyle{ACM-Reference-Format}
\bibliography{references}
\end{document}